\documentclass[10pt,sort&compress]{elsarticle}

\usepackage{graphicx}
\usepackage{textcomp}
\usepackage{natbib}

\usepackage{amsfonts}
\usepackage{amssymb}
\usepackage{amsmath}
\usepackage[usenames]{color}

\usepackage{setspace}
\usepackage{lineno}

\newcommand{\pdr }[2]{\dfrac{\partial {#1}}{\partial {#2}}}

\newcommand{\pddr}[2]{\dfrac{\partial^2 {#1}}{\partial {#2}^2}}

\newcommand{\pdra }[2]{{\partial   {#1}}/{\partial {#2}}}

\newcommand{\tx}{\tilde{x}}

\newcommand{\tr}{\tilde{r}}

\newcommand{\tj}{\tilde{j}}

\newcommand{\tc}{\tilde{c}}

\newcommand{\teta}{\tilde{\eta}}

\newcommand{\tD}{\tilde{D}}
\newcommand{\tN}{\tilde{N}}

\newcommand{\tR}{\tilde{R}}

\newcommand{\cref }{c_{ref}}

\newcommand{\jlim}{j_{GDL}^{lim}}

\newcommand{\veps}{\varepsilon}
\newcommand{\eps}{\epsilon}
\newcommand{\sion}{\sigma_p}

\newcommand{\Dox}{D_{p}}
\newcommand{\tDox}{\tD_{ox}}

\newcommand{\lcat}{l_t}
\newcommand{\jolim}{j_N^{\lim}}
\newcommand{\tjolim}{\tj_N^{\lim}}

\newcommand{\lexp}[1]{\exp\left(#1\right)}
\newcommand{\lnl }[1]{\ln \left(#1\right)}

\newcommand{\etal}{et al.{ }}

\begin{document}

\begin{frontmatter}

\title{The effect of Nafion film on the cathode catalyst layer
       performance in a low--Pt PEM fuel cell}

\author{Andrei Kulikovsky\fnref{label1}}
\ead{A.Kulikovsky@fz-juelich.de}
\fntext[label1]{ISE member}
\address{
Forschungszentrum Juelich GmbH \\
Institute of Energy and Climate Research, \\
IEK--3: Electrochemical Process Engineering \\
D--52425 J\"ulich, Germany
}

\date{\today}

\begin{abstract}
A single--pore model for performance of the cathode catalyst layer (CCL) in a PEM fuel cell
is developed. The model takes into account oxygen transport though the CCL depth and through
the thin Nafion film, separating the pore from Pt/C species. Analytical solution to model
equations reveals the limiting current density $\jolim$ due to oxygen transport
through the Nafion film. Further, $\jolim$ linearly depends of the CCL thickness, i.e., the
thinner the CCL, the lower $\jolim$. This result may explain unexpected lowering of
low--Pt loaded catalyst layers performance, which has been widely discussing in literature.
\end{abstract}


\end{frontmatter}


\section{Introduction}

High cost of platinum could potentially slow down development and marketing
of low--temperature fuel cells. At present, standard oxygen
reduction reaction (ORR) electrode contains 0.4~mg of Pt per square
centimeter. This translates to nearly 100~g of precious metal
in a 100--kW stack for automotive applications.
High Pt cost stimulates worldwide interest in three to four--fold
reduction of Pt loading.

As the cell current density $j_0$ is proportional to Pt surface, one would anticipate
a linear decay of $j_0$ with the Pt loading. However, low--Pt cathodes
exhibit unexpected overlinear performance loss (OPL)~\cite{Grezler_12,Mathias_16}.
So far, the reason for this effect is not fully understood. Grezler \etal\cite{Grezler_12}
attributed poor low--Pt cathode performance to oxygen transport through the Nafion
film covering Pt/C agglomerates. However, to explain the observed transport
resistance, their study has led to unrealistically high Nafion film thickness.
Owejan \etal\cite{Owejan_13} developed a series of cathodes varying Pt loading
and keeping the cathode thickness constant. This has been achieved by diluting
Pt/C by ``pure'' carbon particles. Their results, however, did not allow to make
a definite conclusion about the OPL nature.
Weber and Kusoglu~\cite{Weber_14} discussed the nature of OPL and provided arguments
in favor to the oxygen transport through the Nafion film. Choo \etal\cite{Choo_15} have
found that the OPL in the low--Pt cathode
can partly be mitigated by proper water management of the cell. This work
is another indirect evidence that the origin of the OPL is related to
oxygen transport in Nafion. Kudo \etal\cite{Kudo_16} showed that the dominant part
of the oxygen transport resistivity in the thin Nafion film is due
to Pt/ionomer interface.

Yet, however, the question remains: why oxygen transport in the Nafion film
does not affect the performance of standard electrodes with high Pt loading,
and it has a strong effect on the performance of low--Pt electrodes?
Recent modeling works~\cite{Moore_14,Hao_15,Mashio_17} do not give indisputable
answer to this question.

In this work, we report a single--pore model for the cathode catalyst layer (CCL)
performance. The model
includes oxygen transport in the void pore and in the Nafion film separating
the pore from the Pt/C particles. Analytical solution to model equations shows that
the polarization curve of this system exhibits limiting current density due to
oxygen transport in the Nafion film. Moreover, this limiting current appears to be
proportional to the CCL thickness. This effect may explain OPL of
low--Pt electrodes, as these electrodes are typically
three-- to four times thinner, than the standard Pt/C systems.

\section{Performance equations}

\begin{figure}
	\begin{center}
		\includegraphics[scale=0.8]{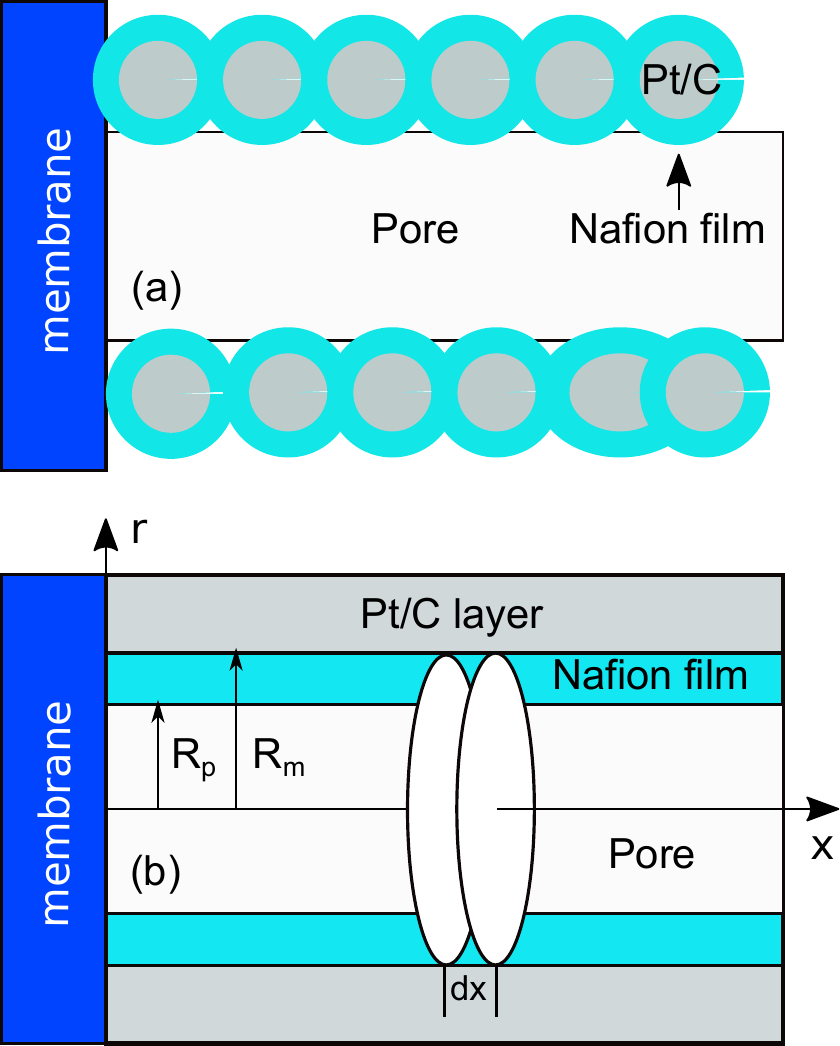}
        	\caption{(a) Schematic of a single pore in the CCL,
                     (b) the system for modeling.
        		}
	    \label{fig:pore}
	\end{center}
\end{figure}

Schematic of a single mesopore in the cathode catalyst layer (CCL) is shown in Figure~\ref{fig:pore}a.
SEM pictures show that in a real CCL, the pore walls are formed by numerous
agglomerates of Pt/C particles, surrounded by a thin Nafion film~\cite{Dobson_12}.
It is assumed that the Pt/C clusters have electric contact between them
for transport of electrons to the ORR sites.

To model this system, we consider a cylindrical pore depicted
in Figure~\ref{fig:pore}b. The system is formed by three coaxial tubes: void,
Nafion film and Pt/C, as shown in Figure~\ref{fig:pore}b.
Let the void pore radius be $R_p$ and the Pt/C radius be $R_m$  (Figure~\ref{fig:pore}).

To write down equations for the pore performance,
we need an oxygen flux balance equation in the pore.
Consider a cylindrical element of the pore volume of the radius $R_m$ and
the thickness $dx$ (Figure~\ref{fig:pore}b). The oxygen diffusive
flux along the pore is $\Dox\pdra{c}{x}$, where $\Dox$ is
the oxygen diffusion coefficient in the pore volume, and $c$ is the local
oxygen concentration in the pore. As $c(x)$ decays toward the membrane,
the balance of fluxes in the volume element is
\begin{equation}
   - \left.\pi R_p^2\Dox\pdr{c}{x}\right|_{x+dx} + \left.\pi R_p^2\Dox\pdr{c}{x}\right|_{x}
      = \left(2\pi R_m dx\right) N_{N,m}
    \label{eq:balance}
\end{equation}
where
\begin{equation}
   N_{N,m} = \left.D_N \pdr{c_N}{r}\right|_{r=R_m}
   \label{eq:NN}
\end{equation}
is the radial oxygen flux at the metal/Nafion interface consumed in the ORR,
$c_N$ is the dissolved oxygen concentration in the Nafion film and
$D_N$ is the oxygen diffusion coefficient in the film. Note
that $D_N$ is the effective parameter, which includes interfacial Nafion resistance.

Dividing both sides of Eq.\eqref{eq:balance} by $\pi R_p^2\, dx$, we come to
\begin{equation}
   - \Dox\pddr{c}{x} = \dfrac{2 R_m N_{N,m}}{R_p^2},
   \quad \left.\pdr{c}{x}\right|_{x=0} =0, \quad c(\lcat) = c_1,
   \label{eq:cx}
\end{equation}
where $c_1$ is the oxygen concentration at the CCL/gas diffusion layer
(GDL) interface, and $\lcat$ is the CCL thickness.

Oxygen transport through the Nafion film along the radial direction is described by
\begin{multline}
    - \dfrac{D_N}{r}\pdr{}{r}\left(r \pdr{c_N}{r}\right) = 0,
    \quad c_N(R_p) = K_H c(x), \\
    \quad \left.D_N\pdr{c_N}{r}\right|_{r=R_m}
        = - \dfrac{i_* R_p^2}{2R_m(4F)}\left(\dfrac{c_{N,m}}{\cref}\right)\lexp{\dfrac{\eta}{b}}
    \label{eq:cNr}
\end{multline}
Here, $K_H$ is the dimensionless Henry constant (mol/mol) for oxygen solubility in Nafion,
$c_{N,m} \equiv c_N(R_m)$,
$i_*$ is the ORR volumetric exchange current density, and
$\cref$ is the reference oxygen concentration.
The first boundary condition in Eq.\eqref{eq:cNr} is obvious, while the second
means that the oxygen flux at the Nafion/Pt interface equals the flux consumed
in the ORR at the Pt surface. This flux is given by the product
of the Tafel ORR rate
$$
  \dfrac{i_*}{4F}\left(\dfrac{c_{N,m}}{\cref}\right)\lexp{\dfrac{\eta}{b}}
$$
by the characteristic length $R_p^2/(2R_m)$. This relation provides correct transition to
the limiting case of vanishing Nafion film. Indeed, setting $K_H =1$, $R_p = R_M$,
we may omit Eq.\eqref{eq:cNr}, set $c_N = c$ and substitute the flux on the
right side of the second boundary condition in Eq.\eqref{eq:cNr} instead of
$N_{N,m}$ into Eq.\eqref{eq:cx}. This gives
\begin{equation}
    - \Dox\pddr{c}{x} = - \dfrac{i_*}{4F}\left(\dfrac{c}{\cref}\right)\lexp{\dfrac{\eta}{b}},
   \label{eq:cxtest}
\end{equation}
which is a standard macrohomogeneous model (MHM) equation for oxygen transport in the CCL.
The system of equations \eqref{eq:cx}, \eqref{eq:cNr}  is completed by the proton
current conservation equation in the Nafion film:
\begin{equation}
  \pdr{j}{x} = - i_* \left(\dfrac{c_{N,m}}{\cref}\right)\lexp{\dfrac{\eta}{b}}
   \label{eq:jx}
\end{equation}

To simplify calculations we introduce dimensionless variables
\begin{equation}
   \tx = \dfrac{x}{\lcat}, \quad \tj = \dfrac{j}{j_p},
                          \quad \teta = \dfrac{\eta}{b}, \quad \tr = \dfrac{r}{\lcat}
   \label{eq:dless}
\end{equation}
where
\begin{equation}
   j_p = \dfrac{\sion b}{\lcat}
   \label{eq:tast}
\end{equation}
is the characteristic current density for proton transport,
$\sion$ is the CCL proton conductivity, and
$b$ is the ORR tafel slope.
With these variables Eqs.\eqref{eq:cx}, \eqref{eq:cNr}
and \eqref{eq:jx} transform to
\begin{equation}
   - \veps_*^2\tDox\pddr{\tc}{\tx} = \eps\veps_*^2\tN_{N,m},
   \quad \left.\pdr{\tc}{\tx}\right|_{\tx=0} =0, \quad
   \tc(1) = \tc_1,
   \label{eq:tc0x}
\end{equation}
\begin{multline}
     \veps_*^2\tD_N\dfrac{1}{\tr}\pdr{}{\tr}\left(\tr \pdr{\tc_N}{\tr}\right) = 0, \quad
    \tc_N(\tR_p) = K_H \tc(\tx), \\
    \left.\eps\veps_*^2\tD_N\pdr{\tc_N}{\tr}\right|_{\tr=\tR_m} = - \tc_{N,m}\exp\teta
    \label{eq:tcN0r}
\end{multline}
\begin{equation}
   \veps_*^2 \pdr{\tj}{\tx} = - \tc_{N,m} \exp\teta
   \label{eq:tj0x}
\end{equation}
Here, $\tc_{N,m}$ and $\teta$ depend parametrically of $\tx$ (see below).

In Eq.\eqref{eq:tc0x}, the dimensionless flux $\tN_{N,m}$ is given by
\begin{equation}
    \tN_{N,m} = \dfrac{4 F\lcat N_{N,m}}{\sion b} = \left.\tD_N\pdr{\tc_N}{\tr}\right|_{\tr=\tR_m},
    \label{eq:tN0r}
\end{equation}
the oxygen diffusion coefficients are normalized according to
\begin{equation}
   \tD = \dfrac{4 F D \cref}{\sion b}
   \label{eq:tD}
\end{equation}
and $\eps$, $\veps_*$ are the dimensionless parameters
\begin{equation}
    \eps = \dfrac{2\tR_m}{\tR_p^2}, \quad \veps_* = \sqrt{\dfrac{\sion b}{i_*\lcat^2}}.
    \label{eq:muveps}
\end{equation}

To find $\tc_{N,m}$ in Eq.\eqref{eq:tj0x} we solve Eq.\eqref{eq:tcN0r}:
\begin{equation}
   \tc_N =  \left(\dfrac{\tR_m\lnl{\tR_m/\tr  }\exp\teta + \eps\veps_*^2\tD_N}
                          {\tR_m\lnl{\tR_m/\tR_p}\exp\teta + \eps\veps_*^2\tD_N}\right)K_H \tc .
   \label{eq:tcN0_sol}
\end{equation}
Setting $\tr =\tR_m$ in this solution, we obtain
\begin{equation}
   \tc_{N,m} = \dfrac{K_H \tc}{1 + \alpha\exp\teta},
   \label{eq:tcN0_sol_Rm}
\end{equation}
where $\alpha > 0$ is a constant parameter:
\begin{equation}
   \alpha = \dfrac{\tR_m}{\eps\veps_*^2\tD_N}\lnl{\dfrac{R_m}{R_p}}.
   \label{eq:alpha}
\end{equation}
With this, Eq.\eqref{eq:tj0x} takes the form
\begin{equation}
   \veps_*^2 \pdr{\tj}{\tx} = - \dfrac{K_H \tc\exp\teta}{1 + \alpha\exp\teta} .
   \label{eq:tj0x2}
\end{equation}
For further references we solve Eq.\eqref{eq:tj0x2} for $\exp\teta$:
\begin{equation}
   \exp\teta = -\dfrac{\veps_*^2\pdra{\tj}{\tx}}{K_H\tc + \alpha\veps_*^2\pdra{\tj}{\tx}} .
   \label{eq:exp0}
\end{equation}


To eliminate $\teta$ from Eq.\eqref{eq:tj0x2}
we differentiate Eq.\eqref{eq:tj0x2} over $\tx$,
substitute $\exp\teta$ from Eq.\eqref{eq:tj0x} and use the Ohm's
law $\tj = -\pdra{\teta}{\tx}$. After simple algebra we come to
\begin{multline}
    \pddr{\tj}{\tx} - \left\{\pdr{\left(\ln\tc\right)}{\tx}
    - \left(1 + \dfrac{\alpha\veps_*^2}{K_H\tc}\,
                     \pdr{\tj}{\tx}\right)\tj \right\}\pdr{\tj}{\tx} = 0, \\
    \tj(0) = \tj_0, \quad \tj(1) = 0
    \label{eq:tj0xfin}
\end{multline}

Eq.\eqref{eq:tcN0_sol} allows us to calculate the flux $\tN_{N,m}$,
which appears in Eq.\eqref{eq:tc0x}.
Calculating derivative $\pdra{\tc_N}{\tr}$, multiplying the result
by $\tD_N$ and setting $\tr = \tR_m$, we get
$$
  \tN_{N,m} = - \dfrac{K_H \tc\exp\teta}{\eps\veps_*^2\left(1 + \alpha\exp\teta\right)}
$$
Using here Eq.\eqref{eq:exp0}, we come to
\begin{equation}
    \tN_{N,m} = \dfrac{1}{\eps}\,\pdr{\tj}{\tx}
    \label{eq:tNp0}
\end{equation}
With this,  Eq.\eqref{eq:tc0x} transforms to
\begin{equation}
   \tDox\pddr{\tc}{\tx}  = - \pdr{\tj}{\tx},
   \quad \left.\pdr{\tc}{\tx}\right|_{\tx=0} = 0,\quad \tc(1) = \tc_1,
   \label{eq:tc0x2}
\end{equation}
Integrating this equation once, we find
\begin{equation}
   \tDox\pdr{\tc}{\tx}  = \tj_0 - \tj, \quad  \tc(1) = \tc_1,
   \label{eq:tc0xfin}
\end{equation}
which is a standard MHM equation for oxygen transport through the CCL depth.

Thus, the problem is reduced to the system of Eqs.\eqref{eq:tj0xfin}, \eqref{eq:tc0xfin}.
With $\tj(\tx)$ and $\tc(\tx)$ at hand,
the overpotential $\teta$ is obtained from Eq.\eqref{eq:exp0}:
\begin{equation}
   \teta = \lnl{-\dfrac{\veps_*^2\pdra{\tj}{\tx}}{K_H\tc + \alpha\veps_*^2\pdra{\tj}{\tx}}}
   \label{eq:teta0_sol}
\end{equation}
and the radial shape of dissolved oxygen concentration $\tc_N$
can be calculated from Eq.\eqref{eq:tcN0_sol}.

\section{Results and discussion}

\begin{figure}
	\begin{center}
		\includegraphics[viewport = 80pt 500pt 350pt 690pt, clip, scale=0.9]{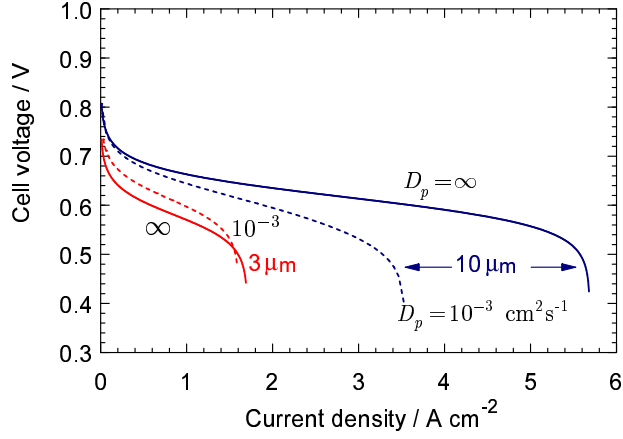}
		\caption{Polarization curves of the catalyst layer of the thickness
         10~$\mu$m and 3~$\mu$m. Solid lines, Eq.\eqref{eq:tvcc} (fast oxygen and proton transport),
         dashed lines, numerical solution to Eqs.\eqref{eq:tj0xfin}, \eqref{eq:tc0xfin}
         (finite rates of proton and oxygen transport).
         All the curves are calculated as $V_{oc} - b\teta$, parameters for the calculation are given
         in Table~\ref{tab:parms}.
         }
		\label{fig:vcc}
	\end{center}
\end{figure}

Consider first the case of fast proton and oxygen transport along the pore.
In that case, the static oxygen concentration $\tc$ and overpotential $\teta$ are
nearly constant through the CCL depth. Integrating Eq.\eqref{eq:tj0x2}
over $\tx$ from 0 to 1, we get
\begin{equation}
   \veps_*^2 \tj_0 = \dfrac{K_H \tc\exp\teta_0}{1 + \alpha\exp\teta_0},
   \label{eq:vcc0}
\end{equation}
where the subscript 0 marks the values at the membrane/CCL interface.
Solving Eq.\eqref{eq:vcc0} for $\teta_0$ we get polarization curve
of the CCL:
\begin{equation}
   \teta_0 = \lnl{\dfrac{\veps_*^2\tj_0}{K_H\tc - \alpha\veps_*^2\tj_0}}
   \label{eq:tvcc}
\end{equation}
Figure~\ref{fig:vcc} shows the polarization curves of the CCL of
the thickness 10~$\mu$m and 3~$\mu$m. Solid lines are calculated using Eq.\eqref{eq:tvcc},
while the dashed lines result from numerical solution of the system
of Eqs.\eqref{eq:tj0xfin}, \eqref{eq:tc0xfin}. The solid lines thus correspond to the
fast rate of proton and oxygen transport in the CCL,
while the dashed lines take into account finite rate of these processes.

As can be seen, all the curves demonstrate limiting current density.
Eq.\eqref{eq:tvcc} helps to understand the effect:
it exhibits the limiting current density $\tjolim$
due to oxygen transport through the Nafion film. This current density
makes the denominator in Eq.\eqref{eq:tvcc} equal to zero.
With $\alpha$ from Eq.\eqref{eq:alpha} and $\eps$ from Eq.\eqref{eq:muveps}, we find
\begin{equation}
   \tjolim = \dfrac{2\tD_N K_H\tc}{\tR_p^2\lnl{\tR_m/\tR_p}} .
   \label{eq:tjlim}
\end{equation}
In the dimension form this equation reads
\begin{equation}
   \jolim = \dfrac{8 F D_N \lcat K_H c}{R_p^2\lnl{R_m/R_p}} .
   \label{eq:jlim}
\end{equation}
In the context of this work, most important is that $\jolim$ is proportional
to $\lcat$, i.e., {\em the limiting current density due to oxygen transport through
the Nafion film linearly decreases with the decrease in the CCL thickness}.
This may explain unexpected poor performance of the low--Pt catalyst layers.
Indeed, lower Pt loading means proportional decrease in the CCL thickness.
For example,
the standard CCL with the Pt loading of 0.4~mg~cm$^{-2}$ is four times thicker,
than the CCL with the Pt loading of 0.1~mg~cm$^{-2}$.
Note that all the other parameters appearing in Eq.\eqref{eq:jlim} are the same
for the thick and thin CCLs.

Suppose that the model above is completed with the oxygen transport
in the GDL. This would lead to another limiting current density $\jlim$ due to
finite oxygen diffusivity of the GDL.
Thus, we may face the situation when in the standard CCL, $\jolim$
exceeds $\jlim$, and the effect of oxygen transport through the Nafion film is not seen.
However, in the low--Pt CCL,  several times lower $\jolim$ may
limit the cell polarization curve.

Qualitatively, the oxygen transport path from the CCL/GDL interface to the Pt
surface consists of two consecutive steps: the transport in the void pore
followed by the transport through the Nafion film. The shorter the pore,
the larger the weight of the transport in Nafion in the overall balance
of oxygen fluxes. This explains proportionality $\jolim \sim \lcat$
in Eq.\eqref{eq:jlim}.

Finite through--plane oxygen diffusion coefficient
$\Dox$ strongly affects the shape of the curve in
the thick 10--$\mu$m CCL, while in the thin 3--$\mu$m CCL,
the effect of $\Dox$ on the limiting
current density is marginal (Figure \ref{fig:vcc}). It is interesting to note
that finite $\Dox$ improves the polarization curve of a thin (low--Pt) CCL in the range
of currents below $\jolim$  (Figure \ref{fig:vcc}). The effect is seemingly
due to redistribution of oxygen concentration along the pore length,
which lowers transport losses in the Nafion film.

To rationalize the dependence of $\jolim$ on the pore radius $R_p$
we will assume that $R_p$ is much larger than the Nafion film thickness.
Thus, $\lnl{R_m/R_p} = \lnl{1 + l_N/R_p} \simeq l_N/R_p$,
where $l_N$ is the Nafion film thickness. Substituting this relation into Eq.\eqref{eq:jlim}
we get
\begin{equation}
   \jolim =  \dfrac{8 F D_N \lcat K_H c}{R_p l_N}
   \label{eq:jlim2}
\end{equation}
Thus, another useful hint from Eq.\eqref{eq:jlim2}
is that $\jolim \sim R_p^{-1}$, i.e., lowering of the mean pore radius
in the CCL increases $\jolim$, making the cell polarization curve less sensitive
to oxygen transport in the Nafion film.
Physically, lowering of $R_p$ means reduction of the total proton current entering
the pore, as another pore takes over part of the current for the conversion.

\begin{table}
\begin{center}
\begin{tabular}{|l|c|}
\hline
Pore radius $R_p$, cm, Ref.\cite{Eikerling_06a}                &   $5\cdot10^{-6}$ (50 nm)    \\
Nafion film thickness $l_N$, cm                               &   $10^{-6}$ (10 nm)    \\
Oxygen diffusion coefficient in                               &   \\
 the Nafion film, $D_N$, cm$^2$~s$^{-1}$, Ref.~\cite{Sethuraman_09} &   $10^{-6}$            \\
Oxygen diffusion coefficient through                          &   \\
 the CCL depth, $\Dox$, cm$^2$~s$^{-1}$, Ref.~\cite{Shen_Knights_11} &   $10^{-3}$            \\
Exchange current density $i_*$, A~cm$^{-3}$                   &   \\
(assumed)                                                     &   $10^{-3}$            \\
ORR Tafel slope $b$, V                                        &   0.03                 \\
CCL proton conductivity $\sion$, $\Omega^{-1}$~cm$^{-1}$      &   0.03                 \\
Henry constant  (mol/mol)                                     &   $6.76\cdot 10^{-3}$    \\
Cell temperature  $T$, K                                      &   $273 + 80$           \\
\hline
\end{tabular}
\end{center}
\caption{Parameters used for calculation of the curves in Figure~\ref{fig:vcc}.
   The ORR Tafel slope and the CCL proton conductivity are taken from impedance
   measurements~\cite{Kulikovsky_16d}.
}
\label{tab:parms}
\end{table}

No attempts to fit the numerical polarization curve following
from Eqs.\eqref{eq:tj0xfin}, \eqref{eq:tc0xfin}
to the experimental curves available in literature have been done. The problem is that
the effective oxygen diffusion coefficient in the CCL $D_{ox}$, which is a composite
containing $\Dox$ and $D_N$ strongly depends on the cell current density~\cite{Kulikovsky_16d}.
A much better alternative to validate the model above is impedance spectroscopy.
This, however, requires development of a transient analog of the model, which will be reported
in a full--length paper.

\section{Conclusions}

A single--pore model for the cathode catalyst layer performance in a PEM fuel cell is developed.
The model takes into account oxygen transport through the CCL depth and through the Nafion film
covering Pt/C agglomerates. In the limit of fast proton and through--plane oxygen transport
in the CCL analytical solution for the CCL polarization curve is derived. This solution reveals
a limiting current density due to proton transport through the Nafion film $\jolim$.
Moreover, $\jolim$ is proportional
to the CCL thickness $\lcat$; thus, in the low--Pt CCL, due to its much lower thickness,
the effect of oxygen transport through the Nafion film may limit the cell current density.
Qualitatively, for a fixed current density $j_0$, the thinner the CCL, the larger
the fraction of oxygen flux
that must be transported through the Nafion film to support $j_0$. Another useful hint from
the analytical result is inverse proportionality of $\jolim$ to the mean mesopore radius $R_p$,
i.e., the CCL with lower $R_p$ is less sensitive to oxygen transport through the Nafion.

\newpage


\clearpage

\centerline{\Large\bf Nomenclature\\[1em]}

\small

\begin{tabular}{ll}
	$\tilde{}$   &  Marks dimensionless variables                             \\
	$b$          &  ORR Tafel slope, V                                        \\
	$c$          &  Oxygen molar concentration in the pore, mol~cm$^{-3}$      \\
	$\cref$      &  Reference oxygen concentration, mol~cm$^{-3}$      \\
	$\Dox$       &  Oxygen diffusion coefficient in the pore, cm$^2$~s$^{-1}$  \\
	$D_N$        &  Oxygen diffusion coefficient in the Nafion film, cm$^2$~s$^{-1}$  \\
	$F$          &  Faraday constant, C~mol$^{-1}$                            \\
	$j$          &  Local proton current density along the pore,  A~cm$^{-2}$     \\
    $\jolim$     &  Limiting current density  \\
                 &  due to oxygen transport in Nafion film, A~cm$^{-2}$     \\
	$j_0$        &  Cell current density, A~cm$^{-2}$                   \\
	$i_*$        &  Volumetric exchange current density, A~cm$^{-3}$          \\
	$\lcat$      &  Catalyst layer thickness, cm                              \\
    $l_N$        &  Nafion film thickness, cm                                 \\
    $R_m$        &  Radius of a Pt/C tube, cm                                \\
    $R_p$        &  Pore radius, cm                                           \\
    $r$          &  Radial coordinate, cm                                     \\
	$x$          &  Coordinate through the CCL, cm                           \\[1em]

\end{tabular}

{\bf Subscripts:\\}

\begin{tabular}{ll}
	$0$      & Membrane/CCL interface \\
	$1$      & CCL/GDL  interface     \\
	$t$      & Catalyst layer         \\[1em]
\end{tabular}

{\bf Greek:\\}

\begin{tabular}{ll}
    $\alpha$            &  Dimensionless parameter, Eq.\eqref{eq:alpha}  \\
    $\veps$             &  Dimensionless Newman's reaction \\
                        &  penetration depth, Eq.\eqref{eq:muveps} \\
    $\eps$              &  $= 2/ \tR_m$   \\
    $\eta$              &  ORR overpotential, positive by convention, V \\
	$\sion$             &  CCL proton  conductivity, $\Omega^{-1}$~cm$^{-1}$  \\
\end{tabular}

\end{document}